\documentclass[twocolumn,preprintnumbers,superscriptaddress,amsmath,amssymb]{revtex4}

\usepackage{amsmath}
\usepackage{graphicx}
\usepackage{dcolumn}
\usepackage{bm}
\usepackage{color}
\usepackage{booktabs, longtable}
\usepackage{multirow}
\usepackage{CJK}
\usepackage{ltxtable}
\usepackage{rotating}
\usepackage[colorlinks, citecolor=red]{hyperref}
\usepackage{verbatim}
\usepackage{amsbsy}
\usepackage{amsfonts}
\usepackage{mathrsfs}
\usepackage[english]{babel}

\begin{document}
	
\title{Neutron skin thickness and its volume and surface contributions in berkelium isotopes}
	
\author{Peng Wang}
\affiliation{School of Physics, Zhengzhou University, Zhengzhou 450001, China}

\author{Zi-Dan Huang}
\affiliation{School of Physics, Zhengzhou University, Zhengzhou 450001, China}

\author{Shuang-Quan Zhang}
\affiliation{State Key Laboratory of Nuclear Physics and Technology, School of Physics, Peking University, Beijing 100871, China}

\author{Ting-Ting Sun}
\email[Contact author,]{ttsunphy@zzu.edu.cn}
\affiliation{School of Physics, Zhengzhou University, Zhengzhou 450001, China}
\affiliation{State Key Laboratory of Heavy Ion Science and Technology, Institute of Modern Physics, Chinese Academy of Sciences, Lanzhou 730000, China}

\begin{abstract}
Accurate determination of the neutron skin thickness ($\Delta R_{\rm np}$) in finite nuclei is essential for constraining the density dependence of the nuclear symmetry energy. This work presents a systematic investigation of $\Delta R_{\rm np}$ for the transuranium berkelium (Bk) isotopes within the framework of the deformed relativistic Hartree-Bogoliubov theory in continuum (DRHBc). The results indicate an overall increase in neutron skin thickness with $N$, which exhibits antikinks at the shell closures $N = 184, 258$ due to the shell effects. A decomposition of $\Delta R_{\rm np}$ into volume and surface terms, based on two-parameter Fermi (2pF) fits to angle-averaged DRHBc densities, demonstrates that the volume term dominates as much as $60\%$--$70\%$ in most nuclei, consistent with the $65\%$ found in $^{208}$Pb, thereby validating the volume-surface decomposition for deformed nuclei and confirming its correlation with the symmetry energy slope $L$. The surface term prevails only near the proton drip line, where the volume fraction drops below $50\%$ due to the reduced neutron-to-proton ratio. Deformation is found to slightly reduce the central radius $R_c$ but markedly enhance the surface diffuseness $a$, leading to a notable increase in $\Delta R_{\rm np}$, primarily driven by the surface term. Furthermore, we extend the decomposition to a directional analysis by extracting 2pF parameters along the symmetry axis ($\theta=0^\circ$) and perpendicular to it ($\theta=90^\circ$). In prolate deformed nuclei, a strong directional dependence is observed: although the nucleus is elongated along the symmetry axis, $\Delta R_{\rm np}$ is significantly larger in the perpendicular direction. This anisotropy is weak for oblate nuclei around the shell closures. The anisotropy of $\Delta R_{\rm np}$ is attributed mainly to the volume term, which remains the dominant contribution in most nuclei regardless of direction. These findings provide new theoretical insights into the interplay among deformation, the neutron skin thickness, and the symmetry energy, and offer a foundation for future experimental probes of neutron skin anisotropy in heavy-ion collisions.
\end{abstract}

\maketitle

\section{Introduction}
\label{Sec_Intro}

The neutron skin thickness, defined as the difference between the root-mean-square (rms) radii of the neutron and proton density distributions, $\Delta R_{\rm np} = r_{\rm n}-r_{\rm p}$, plays an indispensable role in nuclear physics and astrophysics~\cite{EPJA2014Vinas_50_27,NT2023Fang_46_080016,NST2024Ma_35_211, ARNPS2024Mammei_74_321}. It is well established that $\Delta R_{\rm np}$ is strongly correlated with the density dependence of the nuclear symmetry energy $E_{\rm sym}(\rho)$ around the saturation density $\rho_0$ ~\cite{PRL2000Brown_85_5296,NPA2002Furnstahl_706_85,PhysRep2005Steiner_411_325,PRL2005ToddRutel_95_122501,PRL2009Centelles_102_122502}. This correlation is commonly quantified by the slope parameter $L = 3\rho_0 \left. \partial E_{\rm sym}(\rho) / \partial \rho \right|_{\rho_0}$~\cite{PRL2009Centelles_102_122502,PRC2010Carbone_81_041301,PRC2010Chen_82_024321,PRL2009Tsang_102_122701}, which characterizes the stiffness of the symmetry energy. A strong linear correlation between neutron skin in $^{208}$Pb and the slope of symmetry energy, $\Delta R_{\rm np}=0.101+0.00147 L$, was demonstrated in Ref.~\cite{PRL2011RocaMaza_106_252501}. A thicker neutron skin corresponds to a larger value of $L$, indicating higher pressure in neutron-rich matter. Therefore, an accurate determination of the neutron skin thickness in heavy nuclei provides crucial constraints on the density dependence of the nuclear symmetry energy, which has broad implications across multiple domains of physics, such as the structure and reactions of exotic nuclei, the location of nuclear drip lines, nuclear masses, density distributions, collective excitations, and the heavy-ion collision dynamics in nuclear physics~\cite{PhysRep2008Li_464_113,PRC2010Chen_82_024321,PRL2009Centelles_102_122502,AJ2009Xu_697_1549, PRC2016Cai_94_061302, PRC2023Guo_108_034617}, as well as neutron star structure, supernova explosions, neutrino emission, magnetar giant flares, and gravitational wave signals from neutron star mergers in astrophysics~\cite{PRL2009Steiner_103_181101,PRC2009Wen_80_025801,CPC2018Sun_42_025101,PRD2019Sun_99_023004,PRD2023Xia_108_054013}. 

In the laboratory, the radius of the proton density distribution can be determined with high precision by electromagnetic interactions. For instance, the charge radius of $^{208}$Pb has been accurately measured as $r_{\rm ch} = 5.5010(9)$~fm~\cite{ADNDT2004Angeli_87_185}. In contrast, directly and accurately measuring the neutron radius remains challenging. Several experimental approaches have been developed to probe the neutron distribution indirectly, including hadronic scattering~\cite{PRL1981Hoffmann_47_1436,PRC1994Starodubsky_49_2118}, studies of antiprotonic atoms~\cite{PRL2001Trzcinska_87_082501}, parity-violating electron scattering~\cite{NPA1989Donnelly_503_589,PRC1993Horowitz_47_826}, giant dipole resonance~\cite{PRL1991Krasznahorkay_66_1287}, and spin dipole resonances~\cite{PRL1999Krasznahorkay_82_3216}. However, results from these methods often have large uncertainties and tend to be model dependent. For example, analyses based on antiprotonic atom data~\cite{PRL2001Trzcinska_87_082501} typically assume a specific nucleon density shape, such as a two-parameter Fermi distribution, introducing unavoidable systematic uncertainties into the extraction of $\Delta R_{\rm np}$. 

The formation of the neutron skin can arise from either an increase in the volume nuclear radius or an enhancement of the surface diffuseness. The relative significance of these volume and surface contributions is systematically influenced by the stiffness of the nuclear symmetry energy~\cite{PRC2010Warda_81_054309, PRC2010Centelles_82_054314}. Specifically, in the case of a soft symmetry energy ($L \sim 20–60$ MeV), the neutron skin thickness of $^{208}\rm Pb$ comprises nearly equal parts from volume and surface effects; when the symmetry energy becomes very soft ($L < 20$ MeV), the surface contribution becomes dominant, accounting for about $75\%$ of the total thickness; and conversely, in cases with a stiff symmetry energy ($L > 75$ MeV), approximately two-thirds of the neutron skin originates from the volume contribution~\cite{PRC2010Centelles_82_054314}. This systematic relationship underscores the necessity of carefully considering the separation of volume and surface effects when describing neutron skin thickness, as well as incorporating observational constraints when extracting symmetry energy parameters from neutron star properties. Furthermore, the analysis of density distributions for $^{208}$Pb in the PREX-II experiment reveals a pronounced excess of neutrons over protons in both the inner core and the surface region, highlighting the importance of both volume and surface distributions to the symmetry energy in finite nuclei~\cite{PRL2021Adhikari_126_172502}.

As a follow-up investigation of our previous work for transuranium berkelium (Bk) isotopes~\cite{PRC2025Huang_111_034314}, this work focuses on the neutron skin thickness, along with the associated volume and surface contributions. It is noted that Berkelium (Bk, Z = 97) isotopes lie in a key region bridging the known proton shell at $Z=82$ and the predicted magic numbers $Z = 114$, $120$, or $126$, making the study of Bk isotopes crucial for exploring the island of superheavy nuclei. As established in our previous work~\cite{PRC2025Huang_111_034314}, most Bk isotopes have deformation, which reshapes the proton and neutron density distributions, and enlarges the nuclear rms radii significantly. The deformation may also modify the neutron skin thickness, and can even induce an anisotropy in the neutron skin.

Understanding the directional dependence of the neutron skin in deformed nuclei is essential for a complete description of nuclear structure, yet it remains challenging for conventional probes. Traditional experimental methods, such as electron scattering, can only provide angle-averaged radii and thus struggle to directly access such directional information. However, using inelastic $\alpha$-scattering experiments, Ref.~\cite{NPA1994Krasznahorkay_567_521} measured the giant dipole resonance and evaluated neutron skin thicknesses along different directions in $^{150}$Nd. A significantly larger neutron skin thickness was found perpendicular to the symmetry axis ($0.50\pm0.15$~fm) compared to that parallel to it ($0.22\pm0.16$~fm). Although these results do not constitute direct experimental measurements of the anisotropic neutron skin thickness, they still provide important evidence for the possible anisotropy of the neutron skin in deformed nuclei. Besides, in relativistic heavy-ion collisions, by selecting specific collision geometries such as tip-tip and body-body collisions, the anisotropy of nuclear matter distributions becomes experimentally accessible~\cite{PRL2015L_115_222301,PRL2020Giuliano_124_202301}. For example, using zero-degree calorimeters~(ZDCs)~combined with charged-particle multiplicity, distinct collision configurations can be experimentally selected in collisions of deformed nuclei $^{238}$U~\cite{PRL2015L_115_222301}. By correlating the event wise mean transverse momentum with elliptic flow in ultracentral collisions, body-body collision events in deformed nuclei $^{238}$U and $^{129}$Xe have been isolated successfully, offering a new way to extract deformation and neutron skin anisotropy~\cite{PRL2020Giuliano_124_202301}. Furthermore, theoretical studies have shown that measuring the free spectator neutron-to-proton ratio $N_{\rm n}/N_{\rm p}$ in central tip-tip and body-body collisions can reveal the directional dependence of the neutron skin and help constrain the symmetry energy~\cite{NPR2023Liu_40_1,PLBLIU2023_838_137701}. Besides, comparing $N_{\rm n}/N_{\rm p}$ in tip-tip and body-body collisions for $^{96}$Ru+$^{96}$Ru and $^{197}$Au+$^{197}$Au systems allows extraction of the polar angular distribution of the neutron skin, thereby probing the deformed neutron skin. Although selecting events with specific orientations in high-energy collisions with deformed nuclei is very challenging, several promising triggers have been proposed in the literature. The proposed observables can be measured by dedicated zero-degree calorimeters in RHIC (Relativistic Heavy Ion Collider) experiments that have been carried out in recent years~\cite{NPR2023Liu_40_1}. While a direct experimental measurement of the anisotropic neutron skin has not yet been achieved, these studies point to a promising path for future investigation.

Motivated by these deformation correlations, the deformed relativistic Hartree-Bogoliubov theory in continuum (DRHBc)~\cite{PRC2010Zhou_82_011301, PRC2012Li_85_024312, CPL2012Li_29_042101, PRC2020Zhang_102_024314,PRC2022Pan_106_014316,ADNDT2022Zhang_144_101488}, which describes nuclear deformation, pairing correlations, and continuum effects in a unified manner and was used in Ref.~\cite{PRC2025Huang_111_034314}, is also used for the present study. To date, DRHBc theory has achieved remarkable successes in nuclear study, such as the establishment of the nuclear mass table~\cite{PRC2020Zhang_102_024314,PRC2022Pan_106_014316,ADNDT2022Zhang_144_101488,ADNDT2024Guo_158_101661,CSB2021Zhang_66_3561}; shell structure and the  prediction of nuclear magic number~\cite{PRC2023Zhang_107_L041303,PRC2025Huang_111_034314, CPC2024Zheng_48_014107,PRC2024Zhang_110_024302,PRC2025Zhang_112_L051303}; the exploration of the neutron drip line~\cite{IJMPE2021In_30_2150009} and new physics including possible existence of bound nuclei beyond neutron drip lines~\cite{CPC2021He_45_101001, PRC2021Pan_104_024331, PRC2021Zhang_104_L021301} and odd-even differences in the stability “peninsula”~\cite{PRC2024He_110_014301}; the investigation of pairing energy~\cite{PRC2025Mun_111_054305}, one-proton emission~\cite{PLB2023Xiao_845_138160}, and $\alpha$-decay half-lives~\cite{Particles2025Mun_8_42, PRC2024Choi_109_054310}; the description of nuclear charge radii~\cite{PRC2022Kim_105_034340,PRC2023Zhang_108_024310,PLB2023Mun_847_138298,PRC2025Pan_112_024316}; the study of nuclear shapes including shape evolution~\cite{PRC2023Zhang_108_024310, PRC2024Mun_110_024310, PRC2025Huang_111_034314}, shape coexistence~\cite{PRC2022Kim_105_034340, Particles2025Mun_8_32, PRC2025Huang_111_034314, JKPS2020In_77_966}, prolate dominance~\cite{PRC2023Guo_108_014319}, bubble structures~\cite{PRC2022Choi_105_024306}, and inner fission barriers~\cite{CPC2024Zhang_48_104105}; the investigation of halo phenomena, including giant halo~\cite{Particles2024Zhou_7_1128}, deformed two-neutron halo~\cite{PRC2021Sun_103_054315}, nuclear magnetism in the deformed halo nucleus~\cite{PLB2024Pan_855_138792}, and the extension of DRHBc theory to study the shape decoupling effects and rotation of deformed halo~\cite{NPR2024Sun_41_75, SB2021Sun_66_2072}, halos in triaxial nuclei~\cite{PRC2023Zhang_108_L041301}, and unified descriptions of halo nuclei from microscopic structure to reaction ~\cite{PLB2023Zhang_844_138112,PLB2024An_849_138422}. Furthermore, in the DRHBc mass table, systematic investigations for the total binding energies and charge radii of the even-$Z$ nuclei with $8\leq Z \leq 120$ by DRHBc with the PC-PK1 functional~\cite{PRC2010Zhao_82_054319} can reproduce well the available data with a root-mean-square deviation of $\sigma=1.433$~MeV~\cite{PRC2021Zhang_104_L021301,PRC2024Wu_109_024310,AAPPS2025Zhang_35_13}, and $\sigma=0.033$~fm~\cite{ADNDT2024Guo_158_101661}, respectively, demonstrating a great advantage of DRHBc for mass and radius predictions. 
Recent further examinations of to the superheavy nuclear masses and the newly measured masses even highlight the precision of DRHBc mass descriptions~\cite{PRC2021Zhang_104_L021301,PRC2024He_110_014301,NST2025Qu_36_231}.  Besides, the DRHBc calculation for $^{208}$Pb~\cite{PRC2022Kim_105_034340} gives a neutron skin thickness of $\Delta R_{\rm np}= 0.257$~fm, consistent with the PREX-II experimental result of $0.283 \pm 0.071$~fm within uncertainties~\cite{PRL2021Adhikari_126_172502}. All these successful applications demonstrate the reliability of the DRHBc theory.

The present study aims to achieve two main objectives. First, we investigate the relative contributions of the volume and surface terms to the neutron skin thickness and analyze their evolution across the isotopic chain. Second, we examine the possible anisotropy of the neutron skin by comparing its values along the symmetry axis and perpendicular to it. To extract the neutron skin thickness quantitatively~\cite{PRC2010Warda_81_054309}, the Levenberg-Marquardt algorithm is employed to fit the density profiles obtained from the DRHBc calculations to a two-parameter Fermi (2pF) distribution. The skin thickness as well as the volume and surface contributions are then calculated by these 2pF parameters. In order to assess the influence of deformation, results by fitting the densities in the spherical relativistic continuum Hartree-Bogoliubov (RCHB) calculations~\cite{ADNDT2018Xia_121_1} are also provided for comparison. Nevertheless, the present study is performed using the DRHBc theory; other reliable microscopic models including Skyrme, Gogny, and relativistic-mean-field or relativistic-Hartree-Fock functionals, as used in Refs.~\cite{PRC2010Warda_81_054309,PRL2021Adhikari_126_172502}, would also be suitable for such an investigation.

The paper is organized as follows. Section~\ref{Sec_Theo} briefly introduces the DRHBc theory and the formalism for neutron skin thickness along with the volume and surface contributions. After the numerical details in Sec.~\ref{Sec_Numerical}, results and discussions are presented in Sec.~\ref{Sec_Results}. Finally, a summary and perspective are given in Sec.~\ref{Sec_Summary}.

\section{Theoretical framework}
\label{Sec_Theo}

\subsection{DRHBc theory}
\label{subsec:drhb_theory}

Detailed descriptions of the DRHBc theory can be found in Refs.~\cite{PRC2010Zhou_82_011301, PRC2012Li_85_024312, CPL2012Li_29_042101, PRC2020Zhang_102_024314}. Here, for the convenience in discussions, we briefly introduce the formalism. In the DRHBc theory, the relativistic Hartree-Bogoliubov (RHB) equation reads
\begin{equation}
   \begin{pmatrix}
     \hat{h}_D-\lambda &&~\hat{\Delta}\\
      -\hat{\Delta}^* && -\hat{h}_D^*+\lambda
   \end{pmatrix}
   \begin{pmatrix}
      U_k\\
      V_k
   \end{pmatrix}
    =E_k
   \begin{pmatrix}
      U_k\\
      V_k
   \end{pmatrix},
\end{equation}
where $\hat{h}_D$ represents the Dirac Hamiltonian, $\hat{\Delta}$ is the pairing potential, $\lambda$ is the Fermi energy for neutrons or protons, $E_{k}$ is the quasiparticle energy, and $U_k$ and $V_k$ are the quasiparticle wave functions.

The Dirac Hamiltonian in the coordinate space is given by
\begin{equation}
  h_{D}({\bm r})={\bm \alpha}\cdot{\bm p}+V({\bm r})+\beta[M+S({\bm r})],
\end{equation}
where $M$ is the nucleon mass, and $S({\boldsymbol{r}})$ and $V({\boldsymbol{r}})$ are the scalar and vector potentials, respectively. The pairing potential is expressed as
\begin{equation}
 \Delta(\boldsymbol{r}_{1},~\boldsymbol{r}_{2})=V^{pp}\left(\boldsymbol{r}_{1},~\boldsymbol{ r}_{2}\right)\kappa(\boldsymbol{r}_{1},~\boldsymbol{r}_{2}),
\end{equation}
where $\kappa=V^*U^T$ is the pairing tensor and $V^{pp}$ is the pairing force in a density-dependent zero-range type,
\begin{equation}
  V^{pp}\left(\boldsymbol{r}_{1},\boldsymbol{r}_{2}\right)=V_{0}\frac{1}{2}\left(1-P^{\sigma}\right)\delta\left(\boldsymbol{r}_{1}-\boldsymbol{r}_{2}\right)\left(1-\frac{\rho\left(\boldsymbol{r}_{1}\right)}{\rho_{\mathrm{sat}}}\right),
\end{equation}
with $V_0$ being the pairing strength, $\frac{1}{2}(1-P^\sigma)$ the projector for the spin $S=0$ component, and $\rho_{\rm sat}$ the saturation density of nuclear matter.

For an axially deformed nucleus with spatial reflection symmetry, the third component $\Omega$ of the angular momentum $j$ and the parity $\pi$ are conserved quantum numbers. Thus, the RHB Hamiltonian can be decomposed into blocks $\Omega^\mathrm{\pi}$ characterized by $\Omega$ and parity $\pi$. Additionally, the potentials and densities in the DRHBc theory can be expanded in terms of Legendre polynomials~\cite{PRC1987Price_36_354},
\begin{equation}
  f(\boldsymbol{r})=\sum_{\lambda}f_{\lambda}(r)P_{\lambda}(\cos\theta),\quad\lambda=0,2,4,\cdots,
\label{Eq:Legend}
\end{equation}
with
\begin{equation}
  f_\lambda(r)=\frac{2\lambda+1}{4\pi}\int d\Omega f(\boldsymbol{r})P_\lambda(\cos\theta).
\end{equation}

The diagonalization of the RHB matrix gives the quasiparticle wave functions, which can be employed to construct densities and potentials.

The root-mean-square~(rms)~radius is calculated as
\begin{equation}
    r_{\tau}=\langle r^{2}\rangle^{1/2}=\sqrt{\frac{\int d^{3}{\bm r}[r^{2}\rho_{\tau}({\bm r})]}{N_{\tau}}},
    \label{Eq:radii}%
\end{equation}%
where $\tau$ represents the neutron, proton, or nucleon, with $\rho_{\tau}({\bm r})$ and $N_{\tau}$ being the vector density and particle number, respectively. 

The quadrupole deformation is calculated by
\begin{equation}
   \beta_{\tau,2}=\frac{\sqrt{5\pi}Q_{\tau,2}}{3N_{\tau}\langle r_{\tau}^2\rangle},
\end{equation}%
where $Q_{\tau,2}$ is the intrinsic quadrupole moment,
\begin{equation}
    Q_{\tau,2}=\sqrt{\frac{16\pi}{5}}\langle r^{2}Y_{20}(\theta,\phi)\rangle.
\end{equation}

For an odd-$A$ or odd-odd nucleus, the blocking effect of the unpaired nucleon is treated within the DRHBc theory using the equal filling approximation~\cite{CPL2012Li_29_042101, PRC2022Pan_106_014316}.

\subsection{Neutron skin thickness: Volume and surface contributions}
\label{subsec:neutron_skin}

To decompose the neutron skin into surface and volume terms, we adopt the following widely used 2pF distribution for the nuclear density:
\begin{equation}
	\rho_{\rm 2pF}(r)=\frac{\rho_0}{1+\exp[(r-C)/a]},
\label{Eq:rho_2pF}
\end{equation}
where the diffuseness parameter $a$ and the radius $C$ at half-density are determined by fitting densities given by the DRHBc calculations. The fitting is performed using the Levenberg-Marquardt algorithm by minimizing the sum of squared residuals between the two densities, 
\begin{equation}
\chi^{2} = \sum_{i} \left[ \rho_{\rm DRHBc}(r_i) - \rho_{\rm 2pF}(r_i) \right]^{2},%
\end{equation}%
where the sum runs over the discrete radial grid points $r_i$ in the coordinate space.
The central density $\rho_0\approx \rho(0)$ is fixed by the constraint of neutron or proton number,

\begin{equation}
	\int\rho(r)\mathrm{d}^3{\bm r}=\int_{0}^{\infty}\rho(r)4\pi r^{2}\mathrm{d}r=Z, ~~{\rm or}~~ N.
\label{Eq:Ze}
\end{equation}
It is noted that the 2pF densities in Eq.~(\ref{Eq:rho_2pF}) are assumed under spherical symmetry. For deformed nuclei, the parameters are determined by fitting the DRHBc density averaged over angles or along the symmetry axis ($\theta=0^{\circ}$) and perpendicular to it ($\theta=90^{\circ}$). 

Several definitions have been proposed to characterize the nuclear size.
As systematically detailed in the book by Hasse and Myers~\cite{hasse1988geometrical}, the most intuitive way to characterize the nuclear size is by defining the central radius $C$, which coincides with the radius at half density for the 2pF distribution. 

The equivalent sharp radius $R$ is defined as the radius of a sharp distribution with a uniform density corresponding to the nuclear central value $\rho(0)$,
\begin{equation}
 \frac{4}{3}\pi R^3\rho(0)=4\pi\int_0^\infty\rho(r)r^2dr. 
 \label{Eq:R}
\end{equation}

Another equivalent radius is the quadratic radius $Q$, which can be derived directly 
from electron scattering experiments~\cite{ZPhysA1975Suessmann_274_145} and is defined by 
\begin{equation}
  \frac{3}{5}Q^{2}=\langle r^{2}\rangle.
  \label{Eq:Q}
\end{equation}

The surface width $b$ characterizes the diffuseness of the nuclear surface and is defined using the central radius $C$ and the density profile $\rho(r)$. It is derived from the surface weight function
\begin{equation}
g(r) = -\frac{1}{\rho(0)}\frac{d\rho(r)}{dr},
\end{equation}
which is normalized to unity and remains positive for a monotonically decreasing $\rho(r)$. In terms of $g(r)$, the surface width $b$ corresponds to the square root of its second central moment~\cite{ZPhysA1975Suessmann_274_145},
\begin{equation}
b^{2} = \int_{0}^{\infty} (r - C)^{2}  g(r)  dr .
\label{Eq:b_width}
\end{equation}
This integral formulation provides a globally averaged, weighted measure of the surface thickness, reflecting the distribution of the density gradient across the entire nuclear surface.

For a 2pF distribution with diffuseness parameter $a$, the surface width $b$ is related to $a$ by the simple expression
\begin{equation}
  b = \frac{\pi}{\sqrt{3}}a.
  \label{Eq:b}
\end{equation}
In the same context, the surface thickness $t$, defined as the distance over which the density falls from $90\%$ to $10\%$ of its central value $\rho(0)$, is connected to
$b$ through,
\begin{equation}
  t=\frac{4\sqrt{3}\ln 3}{\pi} b \approx 2.42 b,
  \label{Eq:b}
\end{equation}
a relation that holds specifically for Fermi-type distributions~\cite{ZPhysA1975Suessmann_274_145}.

Based on the surface expansion formalism introduced in Refs.~\cite{ZPhysA1975Suessmann_274_145,hasse1988geometrical}, the parameters $C$ and $Q$ can be approximately expressed by the expansions in terms of the surface diffuseness $b$ and the equivalent sharp radius 
$R$, neglecting higher-order terms:
\begin{equation}
	C\simeq R\left(1-\xi^{2}\right),
    \label{Eq:C_expan}
\end{equation}
and
\begin{equation}
   Q\simeq R\left(1+\frac{5}{2}\xi^{2}\right), %
   \label{Eq:Q_expan}%
\end{equation}%
where $\xi=b/R$. Given that $\xi$ is small, typically for thin-skinned nuclei, the higher-order corrections are negligible, and the above approximations remain accurate.

The neutron skin thickness, which measures the radial extension of the neutron distribution beyond proton distribution, is defined as
\begin{equation}
  \Delta R_{\rm np}=\langle r^2_{\rm n}\rangle^{1/2}-\langle r^2_{\rm p}\rangle^{1/2}. 
  \label{Eq:Rnp}
\end{equation}

Finally, from Eqs.~(\ref{Eq:Q}, \ref{Eq:Q_expan}, and \ref{Eq:Rnp}), we obtain a refined expression for the neutron skin thickness~\cite{PRC2010Warda_81_054309},
\begin{equation}
  \Delta R_{\rm np}\simeq\sqrt{\frac{3}{5}}\left[(R_{\rm n}-R_{\rm p})+\frac{5}{2}\left(\frac{b_{\rm n}^2}{R_{\rm n}}-\frac{b_{\rm p}^2}{R_{\rm p}}\right)\right],
\label{Eq:Rnp_2pF} 
\end{equation}
which reveals that the neutron skin thickness arises from two distinct contributions, i.e.,
\begin{equation}
	\Delta R_{\rm np}=\Delta R_{\rm np}^{\mathrm{vol}}+\Delta R_{\rm np}^{\mathrm{surf}}, 
  \label{Eq:Rnp_2pF_2}
\end{equation}
with the surface part defined as
\begin{equation}
  \Delta R_{\rm np}^{\mathrm{surf}}\equiv\sqrt{\frac{3}{5}}\frac{5}{2}\left(\frac{b_{\rm n}^{2}}{R_{\rm n}}-\frac{b_{\rm p}^{2}}{R_{\rm p}}\right),   
  \label{Eq:Rnp_surf}
\end{equation}
and the volume part as
\begin{equation}
	\Delta R_{\rm np}^{\mathrm{vol}}\equiv\sqrt{\frac{3}{5}}(R_{\rm n}-R_{\rm p}).
\label{Eq:Rnp_volume}
\end{equation}

Thus, a neutron skin can form either through a difference in volume radii $R$ between neutrons and protons, or through differences in the surface diffuseness $b$, or a combination of both. The prevalence of these mechanisms in mean-field calculations of finite nuclei will be examined in the following sections.

It may be practical to rewrite the neutron skin thickness as well as the volume and surface contributions in Eqs.~(\ref{Eq:Rnp_2pF}--\ref{Eq:Rnp_volume}) directly in terms of the 2pF parameters in Eq.~(\ref{Eq:rho_2pF}). Using the fact that the diffuseness parameter $a$ in a 2pF function is related to the surface width $b$ by the formula
\begin{equation}
  b = \frac{\pi}{\sqrt{3}}a, 
  \label{Eq:b}
\end{equation}
and inserting relation between the central radius $C$ and the equivalent sharp radius $R$ in Eq.~(\ref{Eq:C_expan}), one can easily get the following expression for the surface and volume terms:
\begin{equation}
\Delta R^{\rm surf}_{\rm np} = \sqrt{\frac{3}{5}}\frac{5\pi^2}{6} \left( \frac{a_{\rm n}^2}{C_{\rm n}} - \frac{a_{\rm p}^2}{C_{\rm p}} \right),
 \label{Eq:Rnp_surf2}
\end{equation}
and
\begin{equation}
 \Delta R^{\rm vol}_{\rm np} = \sqrt{\frac{3}{5}} \left[ (C_{\rm n} - C_{\rm p}) + \frac{\pi^2}{3} \left( \frac{a_{\rm n}^2}{C_{\rm n}} - \frac{a_{\rm p}^2}{C_{\rm p}} \right) \right].
 \label{Eq:Rnp_volume2}
\end{equation}

\section{Numerical Details}
\label{Sec_Numerical}

All numerical details adopted in this work follow those used in the construction of the DRHBc mass tables~\cite{PRC2020Zhang_102_024314,ADNDT2022Zhang_144_101488,PRC2022Pan_106_014316,ADNDT2024Guo_158_101661}. To treat the continuum effects properly, the DRHB equations are solved in the Dirac Woods-Saxon (DWS) basis~\cite{PRC2003Zhou_68_034323,PRC2022Zhang_106_024302}. The following numerical cutoffs and parameters are employed: the angular momentum cutoff in the DWS basis is set to $J_{\max}=\frac{23}{2}h$; the maximum order of expansion in Eq.~(\ref{Eq:Legend}) is $\lambda_{\max}=8$, which has been shown sufficient in previous studies~\cite{IJMPE2019Pan_28_1950082,PRC2022Pan_106_014316}; the box size is set to $20$~fm; and the energy cutoff in the Fermi sea is $E_\mathrm{cut}^+=300$ MeV. For particle-hole channel, the PC-PK1~\cite{PRC2010Zhao_82_054319} density functional is taken. For the particle-particle channel, a zero-range pairing force is used with a saturation density $\rho_{\mathrm{sat}}= 0.152{\rm ~fm}^{-3}$  and a pairing strength $V_0=-325$~MeV$\cdot$fm$^{3}$~\cite{ADNDT2022Zhang_144_101488}.

\section{Results and Discussion}
\label{Sec_Results}

This study systematically investigates the neutron skin thickness in odd-$A$ transuranium Bk isotopes as well as its volume and surface contributions. We first present in {Sec.~\ref{Subsec:A} the neutron skin thickness directly obtained by the self-consistent DRHBc and spherical RCHB calculations. Subsequently in Sec.~\ref{Subsec:B}, the neutron skin thickness is decomposed into volume and surface terms with the 2pF parameters fitted by the angle-averaged DRHBc densities and the deformation effect is quantified by comparing with spherical RCHB results. Finally in Sec.~\ref{Subsec:C}, the anisotropy of the neutron skin thickness has been explored along different directions. 
 
\subsection{Neutron skin thickness of Bk isotopes}
\label{Subsec:A}

\begin{figure}[t!]
\centering
  \includegraphics[width=1.0\linewidth]{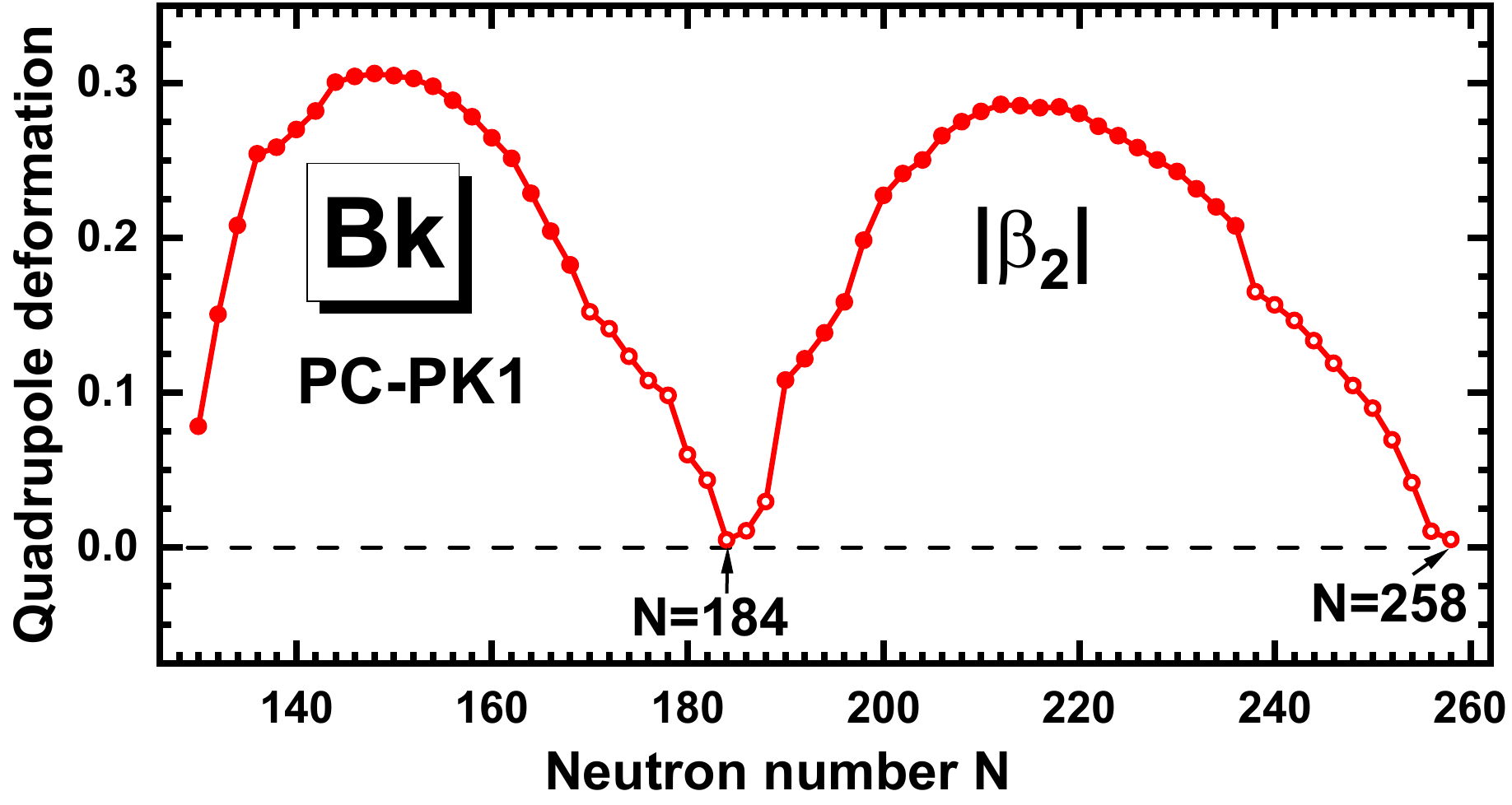}
  \caption{(Color online) Evolution of the quadrupole deformation $|\beta_2|$ for the Bk isotopes in ground states as a function of the neutron number $N$ obtained by the DRHBc calculations with the PC-PK1 density functional. The solid circles denote prolate shapes with positive $\beta_2$ while the open circles denote oblate shapes with negative $\beta_2$.}
\label{Fig:beta2}
\end{figure}

\begin{figure}[t!]
\centering
  \includegraphics[width=0.95\linewidth]{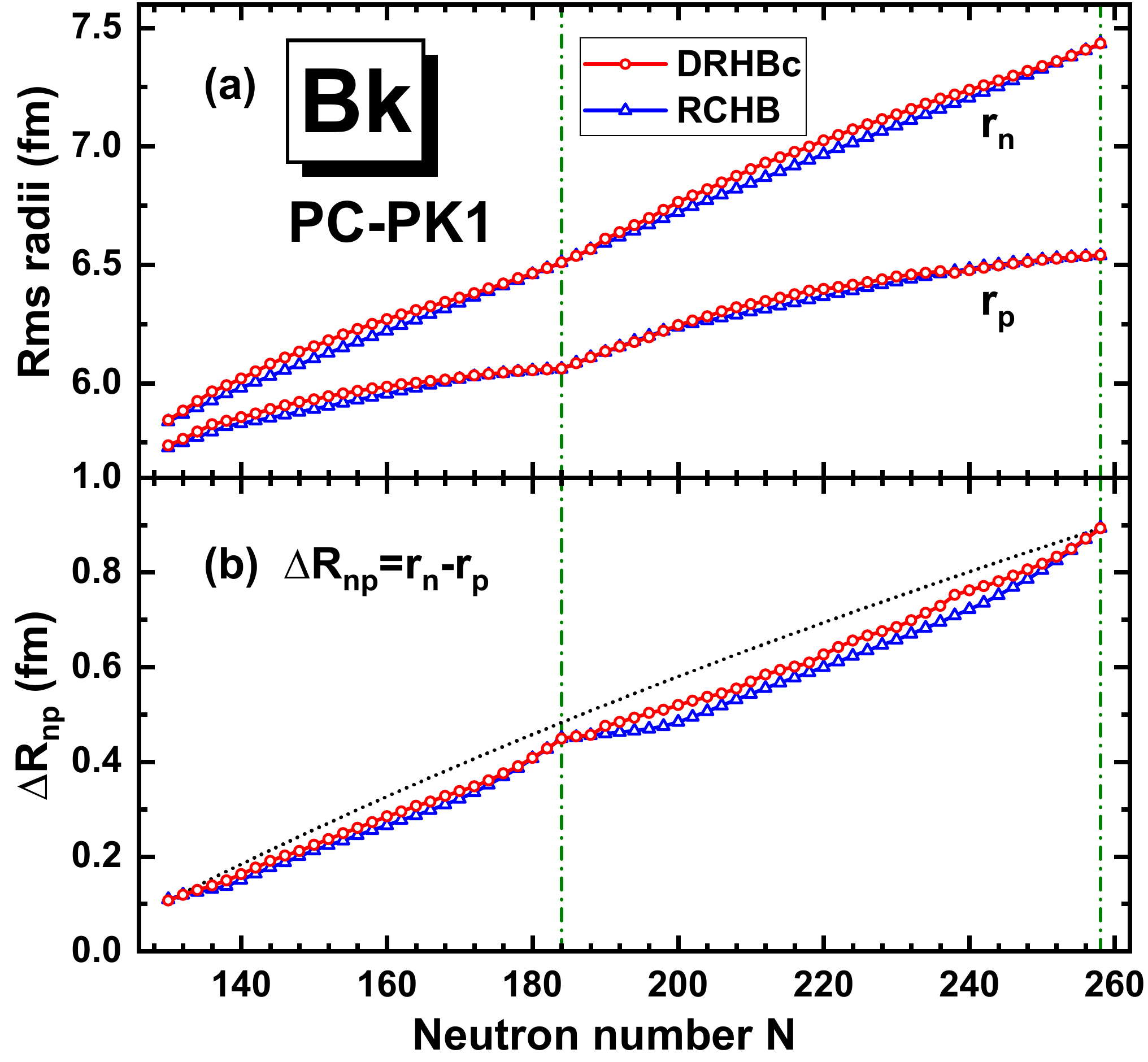}
  \caption{(Color online) (a) Rms radius for neutrons $(r_{\rm n})$ and protons $(r_{\rm p})$, and (b) neutron skin thickness $\Delta R_{\rm np}$, as functions of the neutron number $N$ in Bk isotopes, obtained by self-consistent DRHBc calculations. Spherical RCHB results~\cite{ADNDT2018Xia_121_1} are included for comparison. The dashed line in panel~(b) linking the values of the proton and neutron drip-line nuclei is given to guide the eye.}
\label{Fig:Radii}
\end{figure}

In Fig.~\ref{Fig:beta2}, for the convenience of discussion, we plot the evolution of the quadrupole deformation $|\beta_2|$ for Bk isotopes in the ground states from the proton drip line to the neutron drip line, which are calculated by DRHBc theory with the PC-PK1 density functional. Open and solid circles denote oblate ($\beta_2<0$) and prolate $(\beta_2>0)$ shapes, respectively. Obvious shell structures can be observed with vanishing deformation at neutron closures of $N=184, 258$, while there are pronounced deformations around the mid-shells. Besides, prolate dominance can be found along the whole isotopic chain~\cite{PRC2025Huang_111_034314}.

Next, in Fig.~\ref{Fig:Radii}, the rms radii for neutrons $r_{\rm n}$ and protons $r_{\rm p}$ as well as the neutron skin thickness $\Delta R_{\rm np}=r_{\rm n}-r_{\rm p}$ are presented for the Bk isotopes as functions of neutron number $N$, obtained by DRHBc calculations with the PC-PK1 density functional. The rms radii both for neutrons and protons generally increase with $N$, with noticeable kinks at the neutron magic numbers $N=184$ and $N=258$. Notably, the root-mean-square (rms) neutron radius increases steadily with neutron number both before and after the shell closure at $N=184$, whereas the proton rms radius increases slowly prior to the shell closure but more rapidly afterward, driven by $np$ interactions between valence nucleons and protons. Consequently, these shell-induced differences in the growth rates of neutron and proton rms radii give rise directly to the pronounced antikink structure of neutron skin thickness, despite its overall increasing trend with $N$. Besides, when comparing DRHBc calculations with the spherical RCHB results~\cite{ADNDT2018Xia_121_1}, obvious deformation effect has been suggested to extend the nuclear spatial distributions as well as the neutron skin thickness.

\subsection{Neutron skin thickness averaged over angles: Volume and surface contributions}
\label{Subsec:B}

\begin{figure}[t!]
\centering
  \includegraphics[width=1\linewidth]{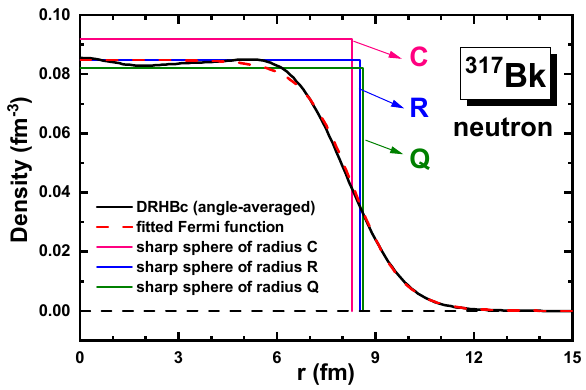}
  \caption{(Color online) Neutron density distribution in $^{317}$Bk. The black solid line denotes the angle-averaged DRHBc result, while the red dashed line represents the corresponding 2pF fit. For comparison, sharp surface density profiles characterized by the central radius $C$, the equivalent sharp radius $R$, and the equivalent rms radius $Q$ are also included.}
\label{Fig:rho_317Bk_averaged}
\end{figure}

To figure out the volume and surface contributions to neutron skin thickness, it turns out that the definition of the nuclear radius must be chosen carefully to avoid misleading interpretations. In Fig.~\ref{Fig:rho_317Bk_averaged}, taking the deformed nucleus $^{317}$Bk as an example, the neutron density distribution obtained by DRHBc and the 2pF fitted profile are displayed. Also shown are sharp-surface densities defined via the central radius $C$, the equivalent sharp radius $R$, and the quadratic radius $Q$. The fitting was performed using the Levenberg-Marquardt algorithm for nonlinear least-squares minimization with the central density constrained by neutron number. As seen in Fig.~\ref{Fig:rho_317Bk_averaged}, the sharp surface density profile with central radius $C$ overestimates the volume density, while that based on quadratic radius $Q$ underestimates it. It is only the profile defined using the equivalent sharp radius $R$ that can reproduce the volume density accurately. Therefore, $R$ is adopted in the following for calculating the neutron skin thickness, as well as the volume and surface contributions.

\begin{figure}[t!]
\centering
  \includegraphics[width=\linewidth]{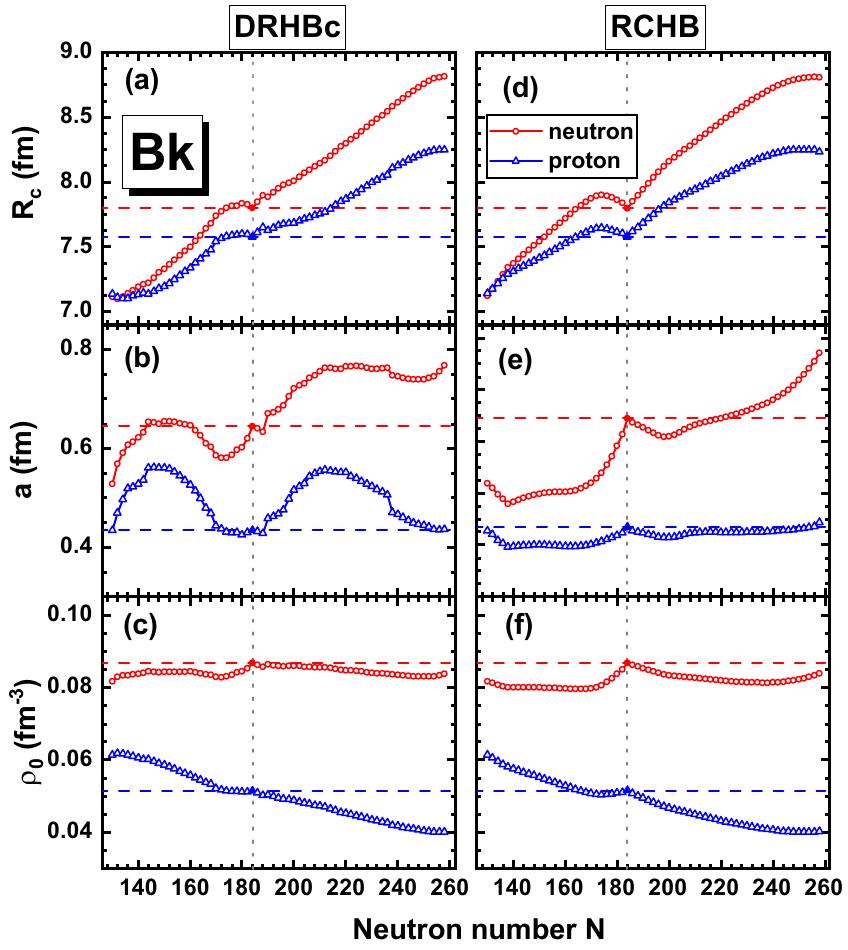}
  \caption{(Color online) Proton and neutron 2pF parameters in Eq.~(\ref{Eq:rho_2pF}) including radius $R_{c}$, diffuseness $a$, and central density $\rho_0$, obtained by fitting the density distributions from (a--c) angle-averaged DRHBc and (d--f) RCHB calculations under charge number constraints. Solid symbols mark the nucleus at the shell closure, with the corresponding horizontal dashed lines indicating its parameter values and the vertical dotted lines denoting the magic number $N=184$.}
\label{Fig:2pF_spherical}
\end{figure}

Figure~\ref{Fig:2pF_spherical} displays the 2pF parameters for protons and neutrons, i.e., half-density radius $R_c~(\equiv C)$, diffusion coefficient $a$, and central density $\rho_0$, extracted from the angle-averaged density distributions by DRHBc [panels a--c] and RCHB [panels d--f] calculations using the PC-PK1 density functional. Solid symbols denote nuclei at the $N = 184$ shell closure, and dashed lines indicate the corresponding reference values of $R_c$, $a$, and $\rho_0$. All 2pF parameters fitted by DRHBc and RCHB densities are consistent well with each other at $N=184$ for both protons and neutrons.

\begin{figure}[t!]
\centering
  \includegraphics[width=0.85\linewidth]{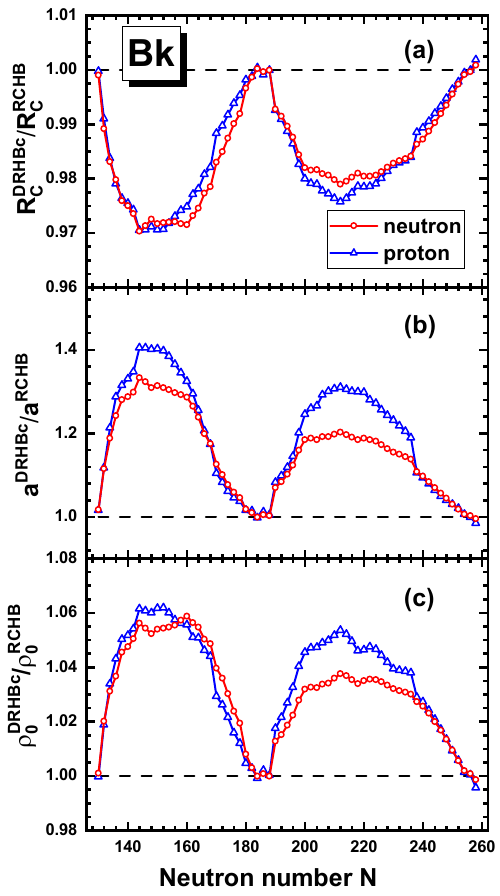}
  \caption{(Color online) Ratios of DRHBc to RCHB results for the (a) radius $R_{c}$, (b) diffuseness $a$, and (c) central density $\rho_0$. Circles and triangles represent results for neutrons and protons, respectively.}
\label{Fig:DRHBcRCHB}
\end{figure}

In panels (a) and (d), the half-density radii $R_c$ obtained by fitting DRHBc and RCHB densities exhibit similar trends in Bk isotopes: a general increase with neutron number $N$ with some localized deviations around the magic number $N = 184$, which is consistent with the kinks observed in the neutron and proton radii in Fig.~\ref{Fig:Radii}(a).
In panels (b) and (e), the evolution of the diffusion coefficients $a$ shows obvious different behaviors between the two theories. In the spherical RCHB calculations, the diffusion coefficient $a$ for neutrons increases with the increasing neutron number $N$, while it remains almost constant for protons. Besides, antikink structures are observed at the neutron shell closure $N=184$ for neutrons, suggesting more diffuse density distributions for closed-shell nuclei. In the deformed DRHBc calculations, besides the effect of more neutrons and the shell effect, nuclear deformation can also have a strong impact and tends to increase diffusion coefficient $a$. For example, the values of $a$ for protons follow trends closely aligned with the deformation evolution shown in Fig.~\ref{Fig:beta2}, while the evolution of $a$ for neutrons is influenced both by the increasing neutrons and deformation. In panels (c) and (f), the volume density $\rho_0$ determined by particle number constraints behaves similarly in both theories: the neutron central density remains nearly constant, while the proton central density gradually decreases due to the increasing proton radius $R_c$ in panels (a) and (d).

To further examine the influence of deformation on the 2pF parameters $R_c$, $a$, and $\rho_0$, Fig.~\ref{Fig:DRHBcRCHB} presents the ratios of the DRHBc values to the RCHB values. Circles and triangles denote the results for neutrons and protons, respectively. In panel (a), when comparing with the quadrupole deformation evolution in Fig.~\ref{Fig:beta2}, we can find that for large deformation, the ratio of central radii $R_c^{\rm DRHBc}/R_c^{\rm RCHB}$ deviates strongly from $1$. This indicates that nuclear quadrupole deformation, of either prolate or oblate shape, can induce a slight contraction of the central radius $R_c$. It is noted that nuclear deformation will increase the nuclear rms radii as shown in Fig.~\ref{Fig:Radii}. Conversely, the diffusion coefficient ratio $a^{\rm DRHBc}/a^{\rm RCHB}$ in panel (b) exhibits the opposite behavior: deformation significantly enhances $a$ for both prolate and oblate deformations, with increases of up to 40\% compared to the spherical case. For the central density $\rho_0$ in panel (c), deformation leads to a slight increase relative to the spherical case. This reflects a radial compression effect in which deformation drives nuclear matter toward the center, thereby increasing the central density. In summary, deformation slightly reduces the central radius $R_c$ and increases the central density $\rho_0$, while significantly enhancing the diffusion coefficient $a$.

\begin{figure}[t!]
\centering
  \includegraphics[width=0.95\linewidth]{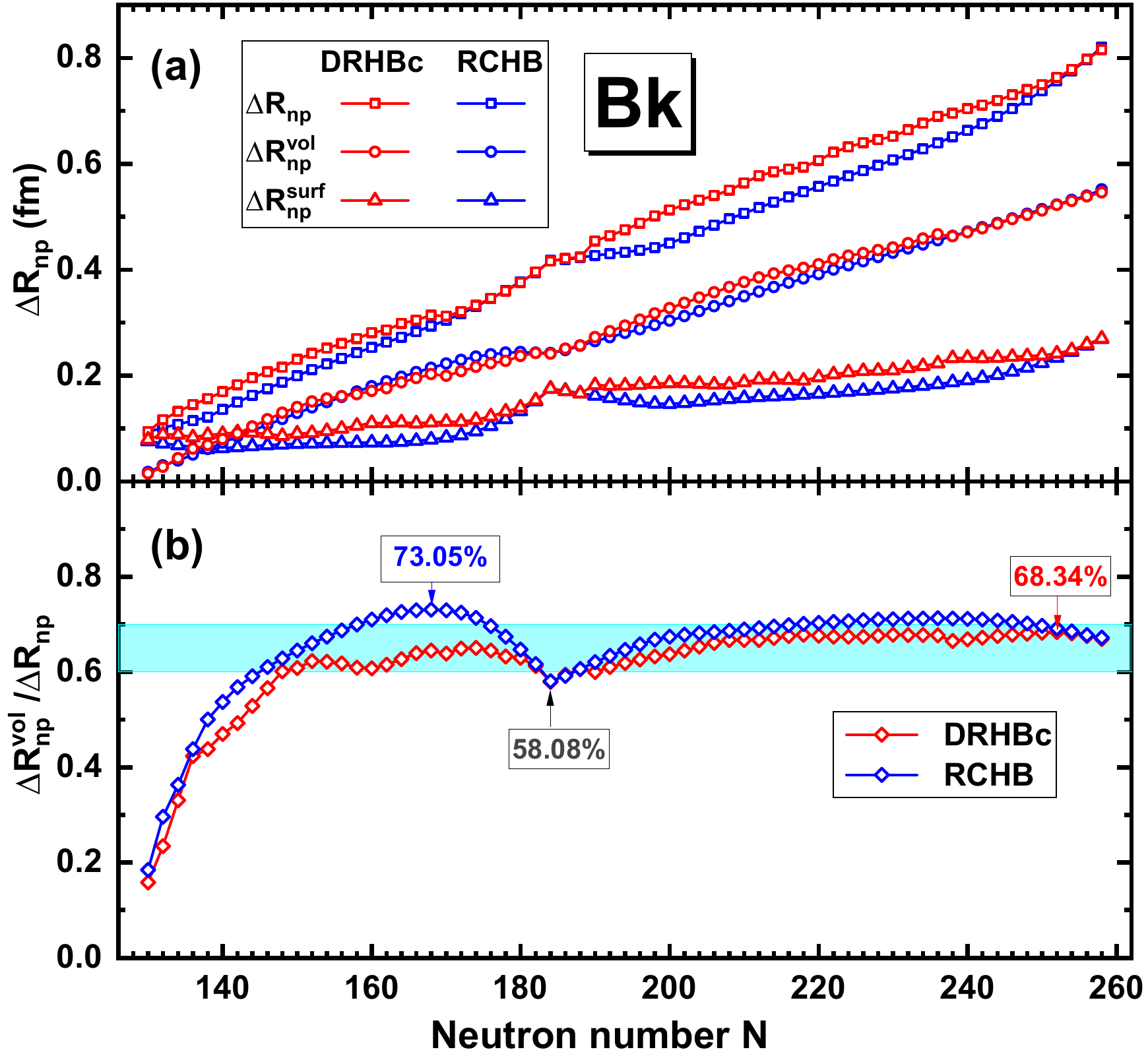}
  \caption{(Color online) (a) Neutron skin thickness $\Delta R_{\rm np}$, along with its volume ($\Delta R_{\rm np}^{\mathrm{vol}}$) and surface ($\Delta R_{\rm np}^{\mathrm{surf}}$) components, and (b) the volume component fraction $\Delta R_{\rm np}^{\mathrm{vol}}/\Delta R_{\rm np}$, as functions of neutron number $N$ in Bk isotopes. The values are obtained using Eqs.~(\ref{Eq:Rnp_surf})–(\ref{Eq:Rnp_volume}) with the two-parameter Fermi (2pF) parameters shown in Fig.~\ref{Fig:2pF_spherical}, which are extracted by fitting the angle-averaged densities from the DRHBc and RCHB models. The cyan region corresponds to a volume component fraction of 60\%--70\%. }
\label{Fig:Rnp_spherical}
\end{figure}

In Fig.~\ref{Fig:Rnp_spherical}, we illustrate neutron skin thickness $\Delta R_{\rm np}$ along with the volume $\Delta R_{\rm np}^{\mathrm{vol}}$ and surface $\Delta R_{\rm np}^{\mathrm{surf}}$ components and the volume component fraction~$\Delta R_{\rm np}^{\mathrm{vol}}/\Delta R_{\rm np}$, as functions of the neutron number $N$ in Bk isotopes calculated by Eqs.~(\ref{Eq:Rnp_2pF}--\ref{Eq:Rnp_volume}), with the 2pF parameters in Fig.~\ref{Fig:2pF_spherical} determined by fitting the DRHBc and RCHB densities. The neutron skin thicknesses extracted from the 2pF functions using Eq.~(\ref{Eq:Rnp_2pF}) have been compared with those directly calculated by the DRHBc theory [see Fig.~\ref{Fig:Radii}(b)], and the differences remain within 4\% for most isotopes, while slightly larger deviations of $7$--$8\%$ appear only around $N = 170-188$ and near the neutron drip line for $N > 238$. In Fig.~\ref{Fig:Rnp_spherical}(a), for spherical nuclei around the magic closures $N = 184$ and $N = 258$, the neutron skin thicknesses predicted by both DRHBc and RCHB are nearly identical. For deformed nuclei, the DRHBc results yield larger neutron skin thicknesses than the spherical RCHB ones, with the enhancement increasing with deformation magnitude, indicating that deformation plays a vital role in determining $\Delta R_{\mathrm{np}}$.

The volume-surface decomposition was originally introduced for spherical nuclei, in particular $^{208}$Pb, to link the neutron skin thickness with the density dependence of the symmetry energy characterized by the slope parameter $L$~\cite{PRC2010Centelles_82_054314}. That decomposition showed that the volume part correlates strongly with $L$, whereas the surface part remains relatively model insensitive. Applying this decomposition to the Bk isotopes in Fig.~\ref{Fig:Rnp_spherical}(b), we find that the volume term dominates in most isotopes, contributing about $60\%$--$70\%$ of the total $\Delta R_{\mathrm{np}}$, a behavior consistent with a moderately stiff symmetry energy. Specifically, the DRHBc results indicate that the volume fraction increases for $N>142$, reaching $68.34\%$ on the neutron-rich side, while the RCHB results give a slightly higher fraction, up to $73.05\%$ at $^{265}$Bk. Notably, for $N<142$ the volume fraction drops below $50\%$, making the surface term dominant due to the reduced neutron-to-proton ratio. The DRHBc calculation with PC-PK1 yields a neutron skin thickness of $0.257$~fm for $^{208}$Pb~\cite{PRC2022Kim_105_034340} and a symmetry energy slope $L=113$~MeV~\cite{PRC2023YNHuang_107_034319}. According to the correlation established in Ref.~\cite{PRC2010Centelles_82_054314}, a volume fraction of about $65\%$ in $^{208}$Pb corresponds to $L=113$~MeV. The $60\%$-$70\%$ volume fraction observed in the Bk isotopes [Fig.~\ref{Fig:Rnp_spherical}(b)] is thus remarkably consistent with that in $^{208}$Pb, validating the volume-surface decomposition for deformed nuclei and suggesting a similar relationship between volume fraction and $L$ even in deformed systems. Nevertheless, unlike $^{208}$Pb for which systematic multi-functional calculations give $\Delta R_{\rm np}=0.101+0.00147L$~\cite{PRL2011RocaMaza_106_252501}, the present work does not yet provide a direct quantitative correlation between $\Delta R_{\rm np}$ and $L$ for the Bk isotopes.

More importantly, by comparing with spherical RCHB results, we find that the deformation-induced enhancement of the neutron skin originates primarily from the surface term. This is explained by the significant increase of the diffusion coefficient $a$ due to deformation [see Fig.~\ref{Fig:DRHBcRCHB}(b)], while the central radii $R_c$ are only slightly reduced [see Fig.~\ref{Fig:DRHBcRCHB}(a)]. These findings demonstrate that the volume-surface decomposition is not only a powerful tool for spherical nuclei but is also equally applicable to deformed nuclei. This work thus provides a solid foundation for extending the volume-surface decomposition to deformed nuclear systems and lays the theoretical groundwork for probing the neutron skin structure of deformed nuclei in future heavy-ion collision experiments.

\subsection{Neutron skin thickness along $\theta=0^{\circ}, 90^{\circ}$: Volume and surface contributions}
\label{Subsec:C}

\begin{figure}[t!]
\centering
  \includegraphics[width=1\linewidth]{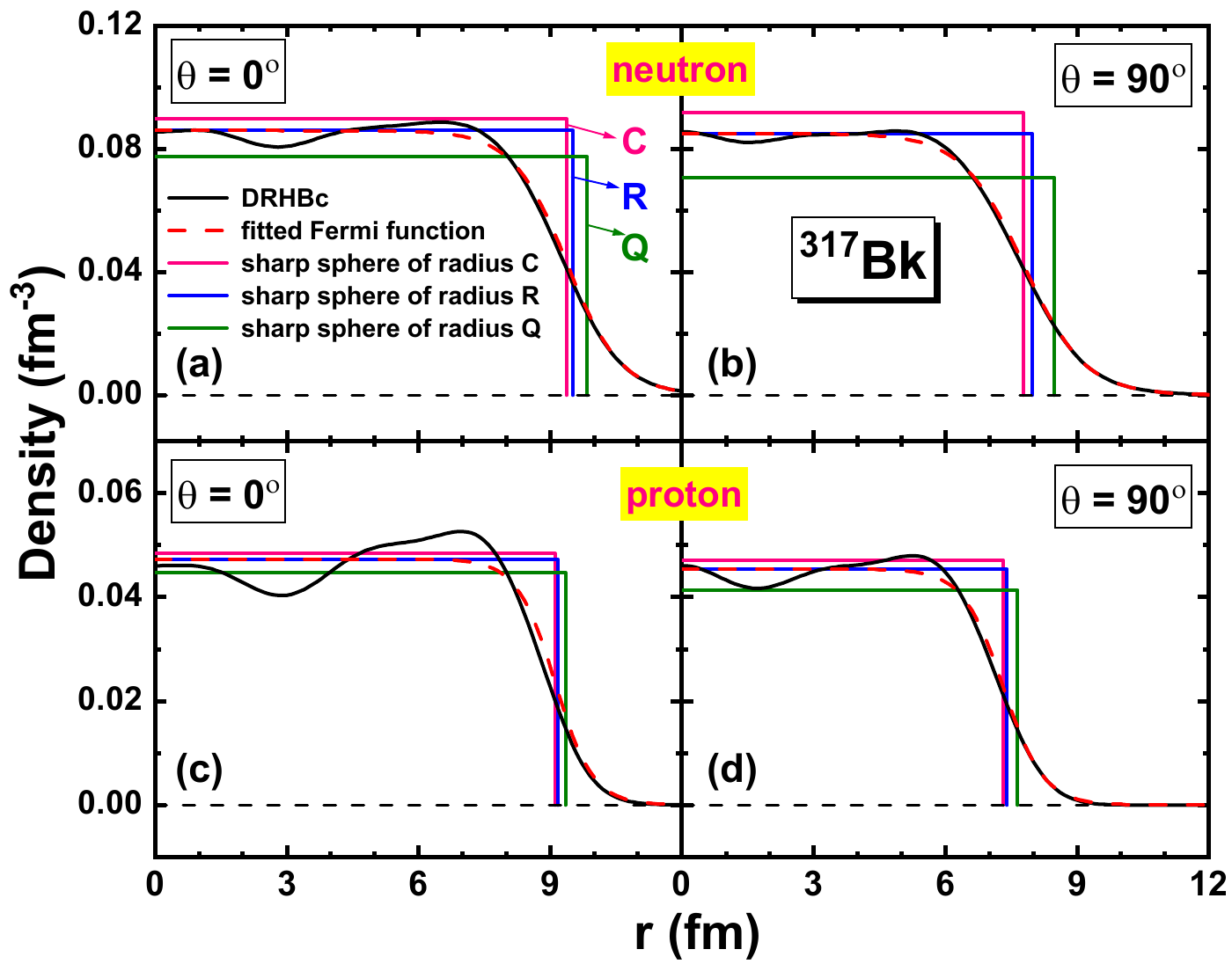}
  \caption{(Color online) Same as Fig.~\ref{Fig:rho_317Bk_averaged}, but showing fits to the DRHBc density distributions along the symmetry axis ($z$, $\theta = 0^\circ$) in (a, c) and perpendicular to that axis ($r_{\perp} = \sqrt{x^2 + y^2}, \theta = 90^\circ$) in (b, d), for neutrons (a, b) and protons (c, d).}
\label{Fig:rho_317Bk}
\end{figure}

In this section, we further investigate the anisotropy of the neutron skin thickness $\Delta R_{\rm np}$ with respect to spatial direction. For this purpose, the 2pF parameters ($R_c$, $a$, and $\rho_0$) are extracted from DRHBc density distributions along the symmetry axis ($z, \theta = 0^\circ$) and perpendicular to it ($r_{\perp} = \sqrt{x^2 + y^2}, \theta = 90^\circ$), respectively. Figure~\ref{Fig:rho_317Bk} displays the density distributions in $^{317}$Bk along $\theta=0^\circ$ (a, c) and $\theta=90^{\circ}$ (b, d) for neutrons (a, b) and protons (c, d). The black solid curves represent the DRHBc results, while the red dashed curves correspond to the 2pF fits obtained via the Levenberg-Marquardt algorithm. It is noted that the density in the internal region exhibits obvious fluctuations, a well-known feature of mean-field densities. In such a case, the 2pF fit remains justified, as the 2pF parametrization is not intended to reproduce these fine details but rather to capture the global behavior of the density distribution, particularly the surface region, which is most relevant for the neutron skin. As shown in Fig.~\ref{Fig:rho_317Bk}, the fitted 2pF function averages over the internal fluctuations and accurately reproduces the surface falloff. Although the DRHBc calculation itself already contains rich physical information and can describe well the density distribution, we still perform the 2pF fit. The purpose is not to replace the self-consistent DRHBc density, but to quantitatively decompose the neutron skin thickness into volume and surface contributions. Therefore, following the approach of Refs.~\cite{PRC2010Warda_81_054309,PRC2010Centelles_82_054314}, we employ the 2pF fit to obtain the key geometric parameters that determine the volume and surface contributions to the neutron skin thickness, and thereby quantitatively decompose the neutron skin thickness into volume and surface parts. For comparison, sharp surface density profiles characterized by the central radius $C$, the equivalent sharp radius $R$, and the quadratic radius $Q$ are also included. In all panels, neither $C$ nor $Q$ accurately reproduces the volume density within the nucleus where $C$ overestimates and $Q$ underestimates the volume density. Only the equivalent sharp radius $R$ matches the volume density correctly. Therefore, $R$ is used to compute the neutron skin thickness via Eqs.~(\ref{Eq:Rnp_2pF}--\ref{Eq:Rnp_volume}).

\begin{figure}[t!]
\centering
  \includegraphics[width=\linewidth]{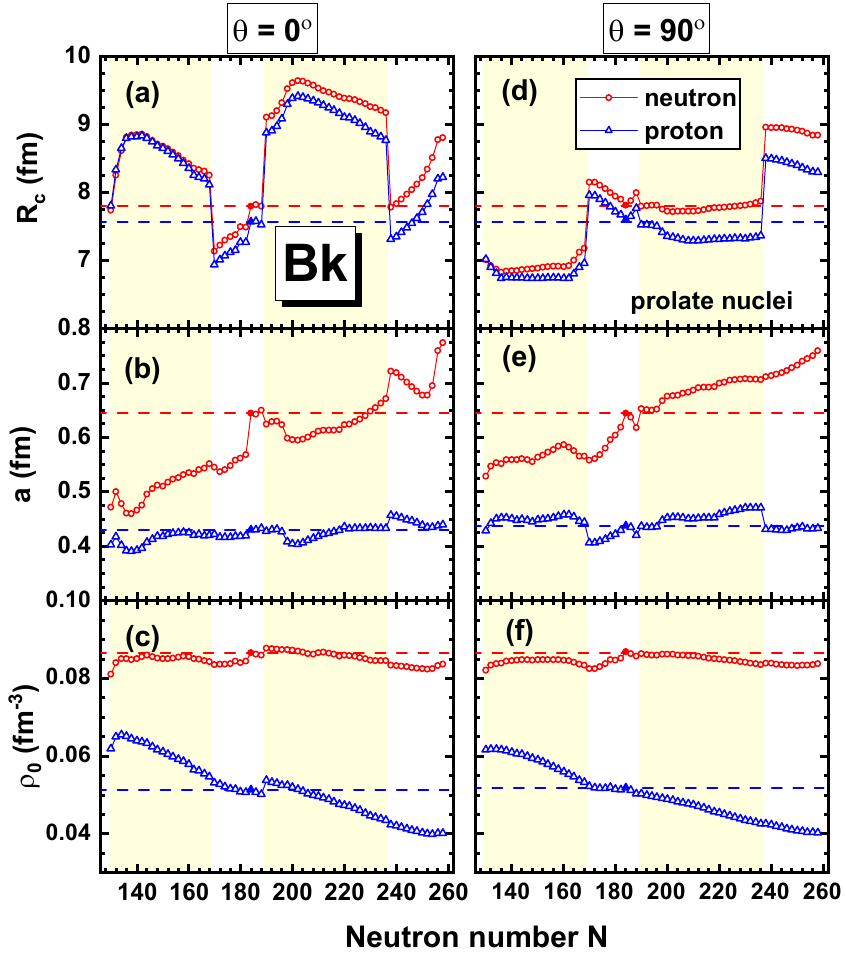}
  \caption{(Color online) Same as Fig.~\ref{Fig:2pF_spherical}, but here fitted to the DRHBc density distributions along the symmetry axis ($z$, $\theta = 0^\circ$) in panels (a--c), and perpendicular to the symmetry axis $(r_{\perp}, \theta = 90^\circ$) in panels (d--f). The yellow region represents nuclei with a prolate deformation.}
\label{Fig:2pF_deformed}
\end{figure}

Figure~\ref{Fig:2pF_deformed} shows the 2pF parameters ($R_c$, $a$, $\rho_0$) extracted from fitting the DRHBc density distributions along the symmetry axis $z$ ($\theta = 0^\circ$) and perpendicular to it ($r_{\perp} = \sqrt{x^2 + y^2}$, $\theta = 90^\circ$), plotted as functions of neutron number $N$ in Bk isotopes. As shown in panels (a) and (d), the central radius $R_{c}$ exhibits strong directional dependence. For prolate nuclei in mass regions $130 \leq N \leq 168$ and $190 \leq N \leq 236$, which are elongated along the symmetry axis $z$, $R_{\rm c}$ is significantly larger along $\theta = 0^\circ$ than along $\theta = 90^\circ$. In contrast, for oblate nuclei in the mass range of $170 \leq N \leq 182$ and $238 \leq N \leq 256$, which are flattened perpendicular to the symmetry axis, $R_{\rm c}$ is larger along $\theta = 90^\circ$ than along $\theta = 0^\circ$. In panels (b) and (e), the diffuseness parameter $a$ exhibits opposite dependence on spatial direction to those for central radius $R_c$, i.e., prolate deformation will increase the diffuseness $a$ along $\theta = 90^\circ$ while oblate deformation increases $a$ along $\theta = 0^\circ$ both for neutrons and protons. In panels (c) and (f), the central densities $\rho_0$, which are determined from particle number constraints along the symmetry axis $z$ and the perpendicular direction defined by $r_{\perp}$ respectively, exhibit no pronounced dependence on spatial direction for both protons and neutrons. Besides, the neutron central density remains approximately constant, whereas the proton central density exhibits an overall decreasing trend with increasing neutron number $N$. 

\begin{figure}
\centering
  \includegraphics[width=0.95\linewidth]{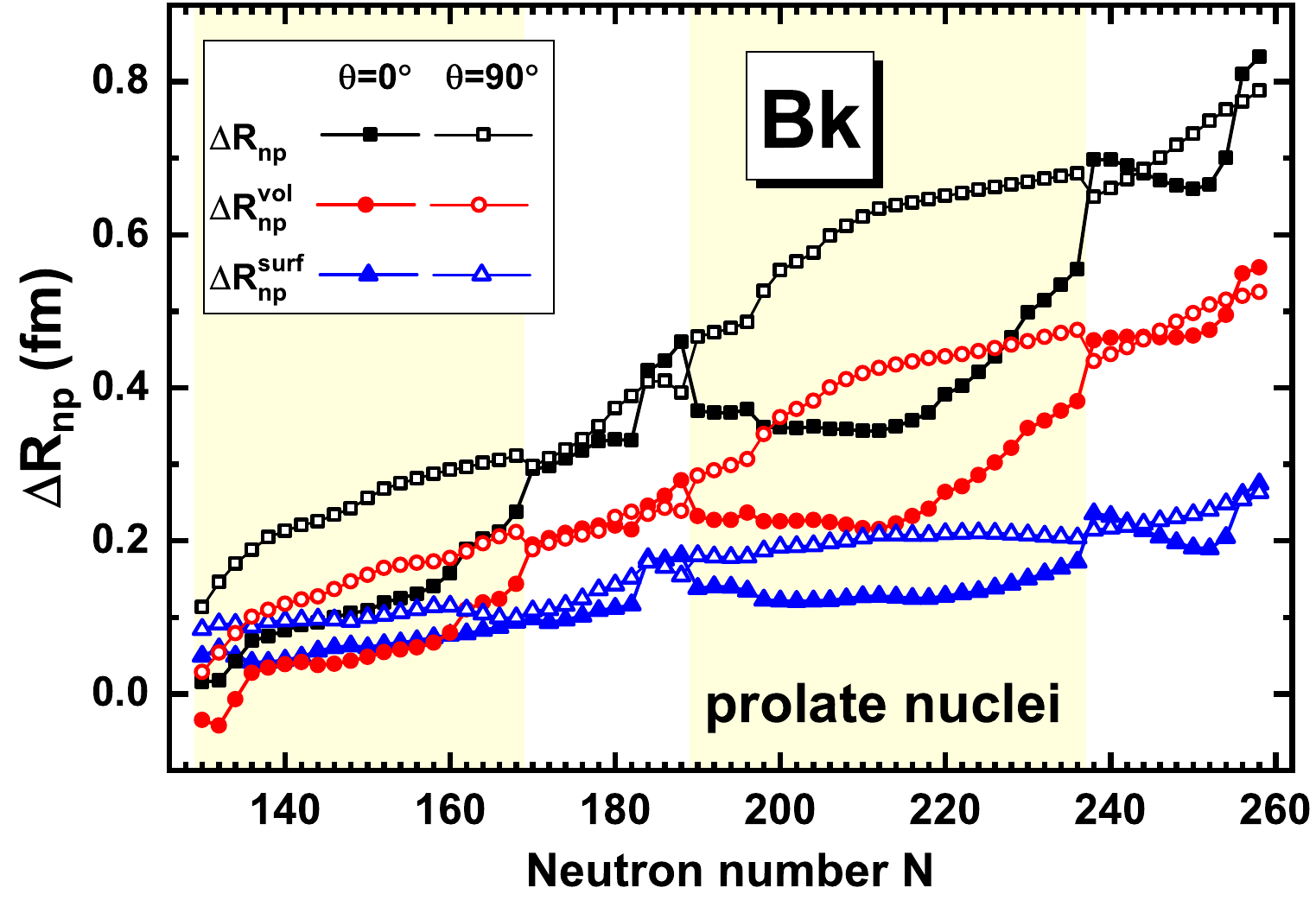}
  \caption{(Color online) Same as Fig.~\ref{Fig:Rnp_spherical}, but showing the neutron skin thickness $\Delta R_{\rm np}$ along with its volume $(\Delta R_{\rm np}^{\mathrm{vol}})$ and surface $(\Delta R_{\rm np}^{\mathrm{surf}})$ components along the symmetry axis ($z$, $\theta = 0^\circ$) and perpendicular to that axis ($r_{\perp}$, $\theta = 90^\circ$). The yellow region represents nuclei with a prolate deformation.}
\label{Fig:Rnpdeform}
\end{figure}

In Fig.~\ref{Fig:Rnpdeform}, we present the neutron skin thickness $\Delta R_{\rm np}$ as well as the volume and surface contributions both along the symmetry axis ($\theta = 0^\circ$) and perpendicular to the symmetry axis ($\theta = 90^\circ$) to study the anisotropy. The 2pF parameters shown in Fig.~\ref{Fig:2pF_deformed} are adopted to do the calculations. A strong directional dependence is observed for neutron skin thickness in prolate deformed nuclei: although they are elongated along the symmetry axis $z$, the $\Delta R_{\rm np}$ perpendicular to the symmetry axis ($\theta = 90^\circ$) is significantly larger than that along the symmetry axis $\theta = 0^\circ$. In contrast, for nuclei in oblate shapes near the magic closures $N=184, 258$, this anisotropy is reduced and the values of $\Delta R_{\rm np}$ along $\theta = 0^\circ$ and $\theta = 90^\circ$ are very close, with a crossing occurring around $N=250$. It is noted that a tiny anisotropy in the neutron skin thickness is still observed for the magic nuclei with $N=184$ and $N=258$, due to their slight nuclear deformation: the quadrupole deformations for neutrons, protons, and the total system are $\beta_2 = -0.003, -0.007, -0.005$ and $\beta_2 = -0.003, -0.010, -0.005$, respectively. When decomposing the $\Delta R_{\rm np}$ into the volume and surface terms, similar anisotropy has been shown in the prolate nuclei. For the volume term, although the central radii $R_{\rm c}$ are larger along $\theta = 0^\circ$ both for neutrons and protons, $\Delta R_{\rm np}^{\rm vol}$ related to $R_{\rm n} - R_{\rm p}$ is smaller along $\theta = 0^\circ$. For the surface term, as the diffuseness $a$ along $\theta = 90^\circ$ is larger than that along $\theta = 0^\circ$ in prolate nuclei, $\Delta R_{\rm np}^{\rm surf}$ related to $b_{\rm n}^2/R_{\rm n}-b_{\rm p}^2/R_{\rm p}$ remains larger along $\theta = 90^\circ$. The crossing of the neutron skin thickness around $N = 250$ between the $\theta = 0^\circ$ and $\theta = 90^\circ$ directions arises from surface effects, driven by the distinct evolution of the neutron and proton diffuseness parameters in oblate nuclei $240 \leq N \leq 258$. As shown in Fig.~\ref{Fig:Rnpdeform}(b), the neutron diffuseness $a_{n}$ first decreases and then increases with neutron number, while the proton diffuseness $a_p$ rises monotonically, and this contrasting behavior leads to the observed crossing. When comparing the contributions from the volume and surface terms, it is found that the volume term dominates $\Delta R_{\rm np}$ as much as $65\%$ in most nuclei with some exceptions near the proton drip line. It is observed that along the symmetry axis $\theta = 0^\circ$ the surface contribution is dominant only for nuclei with $N < 160$, and along the direction $\theta = 90^\circ$ the surface contribution is dominant in nuclei only with $N < 136$. Especially, it is noted that the volume contributions in nuclei with $N=130, 132, 134$ are negative along $\theta = 0^\circ$ as a consequence of $R_{\rm p}$ being slightly larger than $R_{\rm n}$ in those nuclei as shown in Fig.~\ref{Fig:2pF_deformed}(a).

\begin{figure}
\centering
  \includegraphics[width=0.95\linewidth]{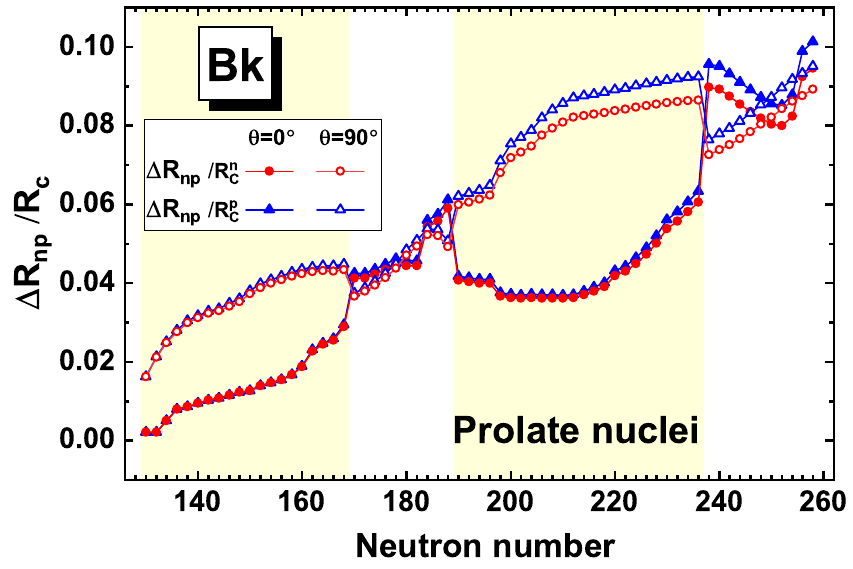}
  \caption{(Color online) Neutron skin thickness $\Delta R_{\rm np}$ along the symmetry axis ($z$, $\theta = 0^\circ$) and perpendicular to it ($r_{\perp}$, $\theta = 90^\circ$), normalized by the central radii of neutrons ($R_{\rm C}^{\rm n}$) and protons ($R_{\rm C}^{\rm p}$). The yellow region denotes nuclei with prolate deformation.}
\label{Fig:Rnp_C}
\end{figure}

To investigate whether the anisotropy of the neutron skin thickness is due to the anisotropy of the neutron (or proton) radii, we normalize $\Delta R_{\rm np}$ along the symmetry axis ($z$, $\theta = 0^\circ$) and perpendicular to it ($r_{\perp}$, $\theta = 90^\circ$) by the central radii of neutrons ($R_{\rm C}^{\rm n}$) and protons ($R_{\rm C}^{\rm p}$), as shown in Fig. \ref{Fig:Rnp_C}. After normalization, $\Delta R_{\rm np}$ still exhibits a clear directional dependence: the value perpendicular to the symmetry axis ($\theta = 90^\circ$) remains significantly larger than that along the axis ($\theta = 0^\circ$). This trend is fully consistent with the anisotropy of the absolute $\Delta R_{\rm np}$ shown in Fig. \ref{Fig:Rnpdeform}, indicating that the directional dependence of the neutron skin thickness is not driven solely by the anisotropy of the neutron (or proton) radii.

\begin{figure}
\centering
  \includegraphics[width=0.95\linewidth]{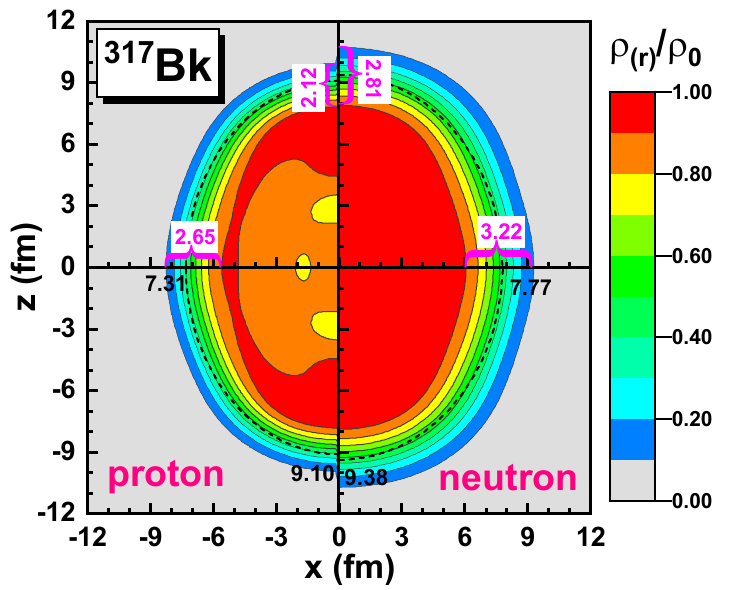}
  \caption{(Color online) Two-dimensional density distributions $\rho(r)/\rho_0$ for neutrons (left) and protons (right) in $^{317}$Bk, plotted in the $x$-$z$ plane. The color scale indicates the density relative to the central density $\rho_0$, ranging from $0\%$ to $100\%$. The surface thickness $t$ (distance for the density to drop from 90\% to 10\% of $\rho_0$) is given for each species along and perpendicular to the symmetry axis $z$. It relates to the surface term of neutron skin thickness $\Delta R_{\rm{np}}^{\rm surf}$ in Eq.~(\ref{Eq:Rnp_surf}) via $t \approx 2.42 b$. Besides, the radii $C_{\rm n}$ and $C_{\rm p}$ at half-density are indicated by dash-dotted lines. These are close to the equivalent sharp radii $R_{\rm n}$ and $R_{\rm p}$, which are associated with the volume term of neutron skin thickness $\Delta R_{\rm{np}}^{\rm vol}$ defined in Eq.~(\ref{Eq:Rnp_volume}).}
\label{Fig:317Bk_rnp}
\end{figure}
Taking the prolate deformed nucleus $^{317}$Bk as an example, Fig.~\ref{Fig:317Bk_rnp} displays the two-dimensional density distribution $\rho(r)/\rho_0$ for protons (left panel) and neutrons (right panel). The color scale indicates the density relative to the central density $\rho_0$, ranging from $0\%$ to $100\%$. To intuitively assess the surface and volume contributions to the neutron skin thickness, we provide the surface thickness $t$, defined as the distance over which the density falls from $90\%$ to $10\%$ of $\rho_0$, and the radii at half-density $C_{\rm n}$ and $C_{\rm p}$ for neutrons and protons, evaluated both along the symmetry axis ($z$) and perpendicular to it ($x$).

For this prolate nucleus, the density distributions of both neutrons and protons extend further along the symmetry axis ($\theta=0^\circ$) than in the perpendicular direction ($\theta=90^\circ$). The radii at half-density along the symmetry axis are  $C_{\rm n}=9.38\;\text{fm}$ and $C_{\rm p}=9.10\;\text{fm}$, significantly larger than those in the perpendicular direction ($C_{\rm n}=7.77\;\text{fm}$ and $C_{\rm p}=7.31\;\text{fm}$). Those values are very close to the values of the corresponding equivalent sharp radii: $R_{\rm n}=9.51\;\text{fm}$ and $R_{\rm p}=9.17\;\text{fm}$ at $\theta=0^\circ$, compared to $R_{\rm n}=7.98\;\text{fm}$ and $R_{\rm p}=7.41\;\text{fm}$ at $\theta=90^\circ$. By comparing $R_{\rm n}$ and $R_{\rm n}$, a larger contribution from the volume term $\Delta R_{\rm np}^{\rm vol}$ is found along $\theta=90^\circ$. Since $\Delta R_{\rm np}^{\rm vol}$ is proportional to $R_{\rm n}-R_{\rm p}$, the measured difference $R_{\rm n}-R_{\rm p}=0.57\;\text{fm}$ in the perpendicular direction exceeds the value of $0.34\;\text{fm}$ along the $z$-axis. This result is consistent with the conclusion drawn from Fig.~\ref{Fig:Rnpdeform}.

To analyze the surface contribution $\Delta R_{\rm np}^{\rm surf}$, we measure the surface thickness $t$, which is related to the diffuseness parameter $b$ via $t \approx 2.42\,b$. Along the $z$ axis, we obtain $t = 2.81\;\text{fm}$ (neutrons) and $2.12\;\text{fm}$ (protons). In the perpendicular direction, although the equivalent sharp radii are smaller, the surface diffuseness are larger, with $t = 3.22\;\text{fm}$ for neutrons and $2.65\;\text{fm}$ for protons. According to Eq.~(\ref{Eq:Rnp_surf}), $\Delta R_{\rm np}^{\rm surf}$ depends on the difference in $b^{2}/R$ between neutrons and protons. Consequently, the larger diffuseness ($t$ or $b$) in the $90^\circ$ direction, combined with the larger equivalent sharp radius $R$ in the $0^\circ$ direction, leads to a larger $\Delta R_{\rm np}^{\rm surf}$ at $\theta=90^\circ$. This result is also consistent with the conclusion drawn from Fig.~\ref{Fig:Rnpdeform}.

Our analysis of neutron skin thickness anisotropy reveals a directional dependence in the spatial separation of neutrons and protons within deformed nuclei. This anisotropy is closely linked to the density dependence of the nuclear symmetry energy. Therefore, the anisotropy uncovered in this work not only advances our understanding of isovector density distributions in deformed nuclei but also provides new theoretical insights and observable signatures for constraining the symmetry energy through geometry-selected measurements in heavy-ion collisions.

\section{SUMMARY and Perspective}
\label{Sec_Summary}

An accurate determination of the neutron skin thickness in finite nuclei provides crucial constraints on the density dependence of the nuclear symmetry energy, which 
further plays an indispensable role in nuclear physics and astrophysics. As the continuation of our previous work for the transuranic Bk isotopes in Ref.~\cite{PRC2025Huang_111_034314}, this work presents a systematic investigation of neutron skin thickness $\Delta R_{\rm np}$ based on the deformed relativistic Hartree-Bogoliubov theory in continuum (DRHBc). For the first time, the volume and surface contributions to $\Delta R_{\rm np}$ are examined for deformed nuclei, and their directional dependence is revealed. 

First, the rms radii for neutrons and protons as well as the neutron skin thickness directly from DRHBc calculations are analyzed. The rms radii both for neutrons and protons generally increase with $N$, with noticeable kinks at the neutron magic numbers $N=184$ and $N=258$. Similarly, the neutron skin thickness also increases with $N$ but exhibits antikinks at neutron shell closures. Deformation notably extends the nuclear spatial distributions and enhances the neutron skin thickness.

Subsequently, the neutron skin thickness is decomposed into volume and surface terms with the 2pF parameters fitted to DRHBc densities averaged over angles. The volume term dominates as much as $60\%$--$70\%$ in most nuclei, consistent with the $\sim\!65\%$ in $^{208}$Pb, validating the decomposition for deformed nuclei and confirming its correlation with the symmetry energy. The surface term prevails only near the proton drip line ($N < 142$), where the volume fraction drops below $50\%$ due to the reduced neutron-to-proton ratio. Deformation slightly reduces the central radius $R_{\rm c}$ but significantly enhances the diffuseness $a$, increasing $\Delta R_{\rm np}$ through the surface term.

Finally, the anisotropy of the neutron skin thickness is explored by extracting 2pF parameters along the symmetry axis ($\theta = 0^\circ$) and perpendicular to it ($\theta = 90^\circ$). In prolate deformed nuclei, $\Delta R_{\rm np}$ is significantly larger in the perpendicular direction, while this anisotropy is weak for oblate nuclei near magic closures. The anisotropy is attributed mainly to the volume term. These findings provide new theoretical insights into the interplay among deformation, the neutron skin thickness, and the symmetry energy, and offer a foundation for future experimental probes of neutron skin anisotropy in heavy-ion collisions. Although angle-averaged radii limit direct experimental access, the 
neutron skin anisotropy is observable through inelastic $\alpha$-scattering (confirmed in $^{150}$Nd) and geometry-selected heavy-ion collisions (via elliptic flow and spectator observables). Future work will apply this analysis to deformed nuclei with measured neutron skin thicknesses, such as $^{150}$Nd.
 
\begin{acknowledgments}	
Helpful discussions with members of the DRHBc Mass Table Collaboration are highly appreciated. This work was partly supported by the Natural Science Foundation of Henan Province (Grant No. 242300421156), the Open Project of the State Key Laboratory of Heavy Ion Science and Technology (Grant No. HIST2026CO04), the National Natural Science Foundation of China (Grants No. 12447101,
No. 12481540215, No. 12435006, and No. U2032141), and the National Key R\&D Program of China (Grant No. 2024YFE0109803).
\end{acknowledgments}

\section*{Data availability}
The data are not publicly available. The data are available from the authors upon reasonable request.


\end{document}